\documentstyle[twoside,fleqn,espcrc2]{article}

\newcommand{\AmS}{{\protect\the\textfont2
  A\kern-.1667em\lower.5ex\hbox{M}\kern-.125emS}}

\title{{\small CPTH-RR469.1096} \\ Non Perturbative Check of N=2, D=4 Heterotic/Type II
  Duality\thanks{Contribution to the proceedings of the {\em Spring
      School and Workshop on String Theory, Gauge Theory and Quantum
      Gravity}, I.C.T.P., Trieste, Italy, March 18-29,
    1996.}\thanks{Research supported in part by the contract
    CHRX-CT93-0340.}}

\author{H. Partouche\address{Centre de Physique Th{\'e}orique, 
        Ecole Polytechnique,
        F-91128 Palaiseau c{\'e}dex, France}}
       
\begin{document}

\begin{abstract}

We review here some of the checks of string-string duality between N=2, D=4
Heterotic and Type II models. The heterotic low energy field theory is reproduced
on the type II side. It is also shown to be a generalization in the string
context of the rigid $N=2$ Yang-Mills theory of Seiberg and Witten which is
exactly known. The non perturbative information of this rigid theory is then
recovered on the type II side. This talk is based on a work done in collaboration with I. Antoniadis \cite{X8}.

\end{abstract}

\maketitle

\section{INTRODUCTION}

We would like here to review some of the arguments for ``String-String Dualities'' \cite{s-s}. Let us first precise what we mean by this. 
Any string theory is defined in perturbation theory. In particular, it is
characterized by a Fock space, whose states have their dynamics dictated by 
string loop calculations. The physics described by such models is perfectly
consistent at any energy scale, but is not complete in the sense that it does
not take into account non perturbative phenomena such as space-time
instantons which are present in field theory. One aim should then be to try
to define at the string level what non perturbative could mean, but this
direction is perhaps out of reach as far as we do not have a second
quantized formulation of string theory. Consequently, each string theory 
should be viewed has the perturbative part of some exact physics. Now, it
would be miraculous if all these supposed exact physics where the same. 
Firstly, this would be a strong sign that with string theory, we go in the
right way to find a unifying complete description of physics. And, by the
way, since all these string perturbation theories are different, they could
describe different perturbative regimes of the underlying exact theory. In
particular, it would become possible to describe non perturbative
effects in the context of a particular string model, simply by studying other
perturbative string models. This equivalence of all these models is the 
so-called {\em String-String Duality}.

Various checks of this scenario are avaible. Some of them, which we review
here,  are explicit and
quantitative; they are based on the following ideas. If one chooses two string theories whose spectra have some
common part, the interactions of this common spectrum must be the same. 
Then, it is consistent to say that the spectra which are not
present in each model are fully non perturbative states. As an exemple,
we will consider here a class of $D=4$, $N=2$ type II and heterotic models \cite{{Vafa},{FHSV}}
which have exatly the same perturbative spectrum. More precisely, we will review the exemple of ref. \cite{X8}; other ones can be found in refs. \cite{{Vafa},{Louis},{Klemm.95},{W2g},{VF}}. Notice that some type I realizations of cousin  models are also given in 
\cite{open}. An important point to note is that $N=2$ space-time supersymmetry implies that the moduli 
space of such a  theory is splitted in a product of two manifolds, one for the
scalars of the vector multiplets and the other for the scalars of the neutral
hypermultiplets. The sigma models based on these manifolds describe the
low energy Lagrangians for the vector multiplets and the neutral hypermultiplets, respectively. All these states are in fact the complete massless spectrum. For type II A compactifications on Calabi-Yau threefolds with
Hodge numbers $h_{11}$ and $h_{12}$, we have $h_{11}$ vector multiplets and the
graviphoton which give rise to the abelian gauge group $U(1)^{h_{11}+1}$. 
Also, there are $h_{12}+1$ neutral hypermultiplets where the $1$ counts for
the type II dilaton. Now, as we said before, the vector multiplets are totaly
decoupled to these hypermultiplets and in particular they do not see at all
the type II coupling constant which sits in the dilaton. Consequently, the
vector multiplets sigma model is not renormalized at all both perturbatively and
non perturbatively. Note that this sigma model
is renormalized in $\alpha'/R$ by world-sheet instantons, where $R$ characterizes the size of the Calabi-Yau. However, the $(2,2)$
underlying superconformal field theory can be realized by a type II B 
compactification on the so-called mirror Calabi-Yau. Consequently, this theory describes the same target space physics. However, the  
vectormultiplet scalars parametrize the complex structure i.e. the shape of this 
mirror manifold and are then independant of its volume. Therefore, their sigma
model can be determined in the mirror Calabi-Yau large volume limit $R^* \rightarrow + \infty$ where $\alpha'/R^*=0$ and
no worldsheet correction survive \cite{{Candelas},{Klemm.93}}. The final type II result obtained by classical
considerations on the world-sheet and string loop expension is
presented in Section 2 in a particular exemple.

In $N=2$ rigid quantum field theory, the analysis of Seiberg and Witten \cite{Seiberg-Witten} showed how the 
knowledge of the singularities on a moduli space which are
interpreted as infrared divergences can be used to determine the full exact 
Lagrangian. Therefore it is natural to extend this idea in the context of strings: if two models have some singularities in common in their respective moduli spaces, we expect that the sigma models based on these manifold should be
the same at least localy around these singularities and therefore the underlying
string theories should be dual. Now, in particular, the type II moduli space
we consider here have a point from which two branches of singularities emerge \cite{Stro}.
This point is similar to the classical singularity
of the $N=2$, $SU(2)$ pure Yang-Mills rigid theory \cite{Seiberg-Witten} where we recover the $SU(2)$
gauge group for zero VEV of the Higgs field. In perturbation theory, this singularity remains and is then splitted in two points when the non perturbative effects are taken into account. If this analogy is valid, a heterotic model which
generalizes this rigid situation has a chance to be dual to the type II original
theory. Such a model can be obtained by a compactification on $K_3 \times T^2$ in order to recover $N=2$ supersymmetry.
Since the heterotic dilaton sits now in one of the vector multiplets, their sigma model has perturbative
and non perturbative corrections. The latter are unknown, and the former will be determined in Section 3 for a particular model,
following the method introduced in \cite{{AFGNT},{deWit}}.
The result will agree with the one found on the type II side and will be the first check of the equivalence
of the two models. 

In Section 4, we will derive the heterotic 
perturbative duality group. By this, we mean the set of T-duality transformations
of the one-loop prepotential (which is in fact the full perturbation theory
due to a $N=2$ non renormalization theorem) and axionic shift. When put in 
matrix representation, we will recover the perturbative monodromy generator
of the rigid Yang-Mills theory. In this context, monodromy means transformation
of an original theory to another wich describes the same physics.

In Section 5, we will find that the perturbative heterotic duality group is
part of the exact type II monodromy group. This will be our second check of
the equivalence of the two theories. Finally, we would like to give a non perturbative
argument for this string-string duality. Even if non perturbative effects are not
defined in the context of string theory, the rigid limit of the low energy
limit of the heterotic model is exactly known. Moreover, this information is totaly
equivalent to the knowledge of the second monodromy generator of the rigid 
theory \cite{Seiberg-Witten}. Therefore, if we identify this matrix in the exact 
type II monodromy group, we will end up with the result that the space-time 
instanton corrections
of the heterotic theory which survive in the flat limit are present in the
type II A side where they originate from world-sheet instanton corrections!
As a final result, we will see that the type II monodromy group contains a 
generator whose interpretation on the heterotic side corresponds roughly 
speaking to an exchange of the coupling constant with the radius of
compactification on the torus.

\section{THE TYPE II MODEL}

Let us consider the rank-3 type II superstring compactification on the Calabi-Yau manifold
$X_{8}$ with Betti numbers $b_{1,1}=2$ and $b_{1,2}=86$, giving
rise to 2 vector multiplets (besides the graviphoton) and $86+1$ hypermultiplets
including the dilaton. This manifold is defined as a hypersurface of degree 8 in
the weighted projective space ${\bf WCP^{4}}[1,1,2,2,2]$. The scalars of the two vector multiplets parametrize the K{\"a}hler structure of $X_{8}$ or,
equivalently, the complex strcture of the mirror $X_{8}^*$ defined as the
zero locus of the polynomial
\begin{equation}
z_{1}^{8}+z_{2}^{8}+z_{3}^{4}+z_{4}^{4}+z_{5}^{4}-8 \psi 
z_{1} z_{2} z_{3} z_{4} z_{5} -2\phi z_{1}^{4} z_{2}^{4} 
\label{X8}
\end{equation}
modded by a discreet group ${\bf Z}_{4}^{3}$. Its complex structure is then 
conveniently described in terms of 
\begin{equation}
x=-\frac{2 \phi}{(8 \psi)^{4}} \quad \mbox{ and }\quad y =
\frac{1}{(2\phi)^{2}}\ .
\end{equation}

The Yukawa couplings are worked out in an explicit form in refs.
\cite{{Candelas},{Klemm.93}}. As follows from $N=2$ special geometry, there
is a preferred coordinate system in which they are given as triple derivatives
of an analytic prepotential \cite{{sugra},{special}}. The corresponding special
coordinates
$t_{1}$ and $t_{2}$ are given by the inverted mirror map, 
\begin{eqnarray}
x(t_{1},t_{2})& =& \frac{1}{h(t_{1})}( 1+ {\cal O}_{t_{1}}(q_{_2}))
\label{mirror map} \\ 
y(t_{1},t_{2})& =& q_{_2} l(t_{1}) ( 1+ {\cal O}_{t_{1}}(q_{_2})) \nonumber
\end{eqnarray}
where $q_{_1}=e^{2i \pi t_{1}}$,  $q_{_2}=e^{2i \pi t_{2}}$ and
${\cal O}_{t_{1}}(q_{_2})$ denotes terms which are at least of order one in
$q_{_2}$  and order zero in $q_{_1}$. The function $l(t_{1})=1+{\cal O}(q_{_1})$,
while $h(t_{1})=1/q_{_1}+104+{\cal O}(q_{_1})$ will be given below. The 
prepotential can then be expressed as
\begin{eqnarray}
{\cal Y} &=& -2t_{1}^{2}t_{2} -\frac{4}{3}t_{1}^{3}+P(t_1,t_2) + g(t_{1}) \label{crochet} \\
  & & +
       g^{^{NP}}(t_{1},t_{2}) \nonumber   
\label{prep II}
\end{eqnarray}
where $P(t_1,t_2)$ is a quadratic polynomial in $t_1$ and $t_2$ whose 
coefficients are integration constants, and 
\begin{eqnarray}
g(t_{1}) &=&{1\over 8i\pi^3}\sum_{n \geq 1}{y_n\over n^3}q_{_1}^{n} 
\label{gdef}\\
g^{^{NP}}(t_{1},t_{2}) &=& \sum_{n \geq 1, m \geq 0}g^{^{NP}}_{nm}
q_{_2}^{n} q_{_1}^{m}\nonumber
\end{eqnarray}
for $y_n$ and $g^{^{NP}}_{nm}$ known integer coefficients related to world-sheet instanton numbers.


The above form of the prepotential suggests that this type II model has a
heterotic dual realization by identifying $t_2$ with the heterotic
dilaton $S$ and $t_1$ with the single heterotic modulus $T$ 
\cite{Vafa}. In fact, the general rank-3 heterotic prepotential $F$ takes the form
\begin{equation}
{ F} = -2ST^{2} + f(T) + f^{^{NP}}(T,S)\ ,
\label{prep}
\end{equation}
where the first two terms of the right hand side correspond to the classical and
one-loop contributions, respectively. $f^{^{NP}}$
denotes the non perturbative corrections
\begin{equation}
f^{^{NP}}(T,S) = \sum_{n \geq 1}f^{^{NP}}_{n}(T)
q_{_S}^{n}
\end{equation}
where $q_{_S}=e^{2i\pi S}$ and $\langle
S\rangle =\frac{\theta}{\pi}+i \frac{8\pi}{g_s^2}$ in terms of the
$\theta$-angle and the four-dimensional string coupling constant $g_s$.

This identification is motivated by the fact that in the weak coupling limit
$q_{_2}\rightarrow 0$ (or equivalently $y\rightarrow 0$), the discriminant locus
of the mirror Calabi-Yau $X_{8}^{*}$
\begin{equation}
\label{conifold}
\Delta=\{ (1-2^{8}x)^{2}-2^{18}x^2y\}\{ 1-4y\}=0
\end{equation}
becomes a perfect square, in complete analogy with the situation in the
rigid $SU(2)$ $N=2$ supersymmetric Yang-Mills theory where the classical
singular $SU(2)$ point splits into two separate branches in the quantum theory
\cite{Seiberg-Witten}. The first factor in eq. (\ref{conifold}) defines
the conifold singularity while the second factor gives rise to an isolated
singularity $y=1/4$ corresponding to the infinite strong coupling limit
$S\rightarrow 0$ in the heterotic theory \cite{{Candelas},{Vafa}}.

The guide to construct the heterotic realization is to first find the classical
$T$-duality group \cite{{Klemm.95},{Yau}}. This can be determined by the set of
transformations which identify all solutions of the conifold
singularity in the weak coupling limit $y=0,x=1/256$ in terms of the special
coordinate $T$,
\begin{equation}
h(T)=256 \quad\mbox{ and }\quad S=i\infty\ ,
\label{h=256}
\end{equation}
where $h$ is defined in eq. (\ref{mirror map}). In fact $h$ was found to be (up
to an additive constant 104) the Haupmodul for ${\Gamma_{0}(2)}_+$ which should
play the role of the heterotic $T$-duality group,
\begin{equation}
h(T) = \frac{\left[ \, \theta_{3}^{4}(T)+\theta_{4}^{4}(T) \,
\right]^{4}}{16\left[ \,  \eta(T)\eta(2T) \, \right]^{8}}\ ,
\label{hfun}
\end{equation}
where $\theta_{3,4}$ are the Jacobi $\theta$-functions and $\eta$ the Dedekind
function.

${\Gamma_{0}(2)}_+$ is the group of modular transformations \cite{Conway}
\begin{equation}
T \rightarrow \frac{aT+b}{cT+d} \equiv MT\quad ;\quad M= 
\left( \begin{array}{cc}
         a & b \\
         c & d
       \end{array} \right)
\label{Mt}
\end{equation}
with $a,b,c,d$ integers, generated by any two of the transformations 
\begin{eqnarray}
W &:& T\rightarrow -\frac{1}{2T} \ , \nonumber\\
U &:& T \rightarrow T+1 \ , \label{W} \\
V &:& T \rightarrow -\frac{2T+1}{2T} \nonumber \ .
\end{eqnarray}
$W$ and $V$ generate the little groups of the fixed points $i/\sqrt 2$ 
and $(i-1)/2$ which are of order 2 and 4, respectively. $h$ is a bijection
between the $\Gamma_0(2)_+$ fundamental domain  and the Riemann sphere with
three punctures such that $h(i/\sqrt 2)=256$, $h((i-1)/2)=0$ and
$h(i\infty)=\infty$. Finally the solutions of the equation (\ref{h=256}) for
the conifold singularity in the perturbative limit consist in the images of
$i/\sqrt 2$ under the action of ${\Gamma_{0}(2)}_+$, which shows also that this
group is the modular one for $T$. 

The dual of the above type II model should be a rank-3 $N=2$ heterotic vacuum
with one vector multiplet modulus $T$ besides the dilaton $S$, and 87
hypermultiplets. The method of constructing such models is described in
ref. \cite{Vafa}, although the exact $K_3$ compactification which leads to
this particular spectrum is not presently known. Here we concentrate on the
$T^2$ lattice part which is relevant for understanding the moduli space of
vector multiplets. A heterotic dual should have a one-loop correction to
its prepotential (\ref{prep}) $f(T)$ equal to the type II quantity
$g(t_1)$ which originates from world-sheet instantons (\ref{prep II}).
In order to determine such a one-loop contribution $f(T)$ which is actually the complete
perturbative correction due to a $N=2$ non renormalization theorem based on analyticity and the invariance under the axionic shift. In $N=2$ supergravity, the K{\"a}hler metric takes the form 
\begin{equation}
{ K}_{T {\bar T}} = K^{(0)}_{T \bar{T}}\left[1+
\frac{2i}{S- \bar{S}}{\cal I} + \cdots\right]\ ,
\label{K_ttbar}
\end{equation}
where the tree-level metric $K^{(0)}_{T\bar{T}} =-2/(T-\bar{T})$ and
${\cal I}(T,{\bar T})$ is given by
\begin{equation}
{\cal I} =\frac{i}{8} \left( \partial_T-\frac{2}{T-
\bar{T}} \right)
 \left( \partial_T-\frac{4}{T- \bar{T}} \right) f + c.c.
\label{I}
\end{equation}
Now, ${\cal I}$ can be computed by a one loop string calculation of an amplitude
involving the antisymmetric tensor, using the method of ref. \cite{yuka}. 
The result must be invariant under modular transformations on $T$. From
the expression (\ref{I}), one can easily deduce the fifth derivative of the one
loop prepotential, $f^{(5)}$, and see that it is a modular function of
weight 6. An expression for $f(T)$ up to a quartic
polynomial is then
\begin{equation}
f(T)= \int_{T_{0}}^{T} \frac{(T-T')^{4}}{4!} f^{(5)}(T') dT'\ ,
\label{forme integrale}
\end{equation}
where $T_0$ is an arbitrary point. The path of integration should not cross any
singularity of $f^{(5)}$, while the result of the integral depends on the
homology class of such paths. Different choices of homology classes of
paths change $f$ by quartic polynomials. Moreover under a modular
transformation (\ref{Mt}), $f$ does not transform covariantly. Using its
integral representation (\ref{forme integrale}), we see that it has a
weight $-4$ up to an addition of a quartic polynomial, ${\cal M}(T)$,
\begin{equation}
f(MT)=(\det M )^{2} (cT+d)^{-4} [f(T)+{\cal M}(T)]\ .
\label{transfo modulaire}
\end{equation}
In order to leave the physical metric invariant, one can see that the
polynomials ${\cal M}(T)$ are quartic with real coefficients.

Finally, in order to guarantee modular invariance of the full effective action,
the dilaton should also transform. Imposing the general condition that
modular transformations may be compensated by K{\"a}hler transformations, one has to redifine the dilaton $S$ to  
\begin{equation}
S -{c\over 2}\frac{f_T + {\cal M}_T}{cT+d} +c^2
\frac{f+{\cal M}}{(cT+d)^2} +\frac{{\cal M}_{TT}}{12}+\lambda_M ,
\label{Stransf}
\end{equation}
where the additive real constant $\lambda_M$ correspond to the
perturbative
heterotic symmetry of the axion shift.

The above discussion applies to any rank-3 $N=2$ heterotic string
compactification. We now specialize to the dual candidate of the type II model
described in Section 2, for which the classical duality group is
${\Gamma_{0}(2)}_+$. Furthermore, the order 2 fixed point in its fundamental
domain, $T=i/{\sqrt 2}$, which was found from the conifold
singularity in the type II model, should correspond in the heterotic
realization to the perturbative $SU(2)$ enhanced symmetry point. Imposing
these two constraints, it is possible to find an explicit heterotic model \cite{X8}. Its
one-loop K{\"a}hler metric involves 
\begin{eqnarray}
{\cal I}  &=& \sum_{\ell=1}^6 \int\frac{d^2\tau}{\tau_{2}^{\; 3/2}}
\; {\bar C}_{\ell}(\bar{\tau}) \; \partial_{\bar{\tau}}  \left(\tau_2^{\; 1/2} Z \right) \\
Z &=& \sum_{p_L,p_R\in \Gamma_{\ell}}e^{\pi i\tau\, |p_{L}|^2} \,  
e^{-\pi i \bar{\tau\,} p_{R}^2}   \ ,   \nonumber               
\label{I1}
\end{eqnarray}
where the sum over $l$ extends over six lattices denoted by $\Gamma_{\ell}$ which parametrise the left and right momenta
\begin{eqnarray}
p_L &=& \frac{i\sqrt{2}}{T-\bar{T}} (n_1 + n_2 \bar{T}^2 + 2m
\bar{T} )
\nonumber \\
p_R &=& \frac{i\sqrt{2}}{T-\bar{T}} (n_1 + n_2 T\bar{T} + 
m (T+\bar{T}) )\, .
\label{mom}
\end{eqnarray}
$\bar{C}_{\ell}$'s are $T$-independent modular functions with well-defined
transformation properties dictated by modular invariance of the integrand.
Actually $\bar{C}_{\ell}$ is the trace of 
$(-1)^F q^{L_0-c/24} \bar{q}^{\bar{L}_0-\bar{c}/24}/{\bar\eta}$ in the Ramond
sectors of the corresponding remaining conformal blocks. Finaly, the integral
is over the Teichm{\"u}ler parameter of the world-sheet torus inside its
fundamental domain. A corresponding expression of the fifth derivative of
the one loop propotential $f^{(5)}(T)$ can then be obtained. A study of its
singularities and the knowledge of its weight six under $\Gamma_0(2)_+$ modular transformations can be used to get the expression
\begin{equation}
f^{(5)}(T) = \frac{64}{i\pi} \left({h_T\over h-h\left({i\over{\sqrt 2}}\right)}\right)^3
{5h+3h\left({i\over{\sqrt 2}}\right)\over h^2} \ .
\label{d5f}
\end{equation}

A first non-trivial check of the proposed duality between this heterotic
model and
the type II compactification described in Section 2, is the comparison of
the two
corresponding prepotentials $F$ and ${\cal Y}$ of eqs. (\ref{prep}) and 
(\ref{crochet}). Indeed, the identification $t_1=T$ and $t_2=S$ should imply that
\begin{equation}
f^{(5)}=g^{(5)}=4\pi^2\sum_{n \geq 1}n^2 y_nq_{_{1}}^{n} \ ,
\end{equation}
which we verified up to the fifth order in the $q_{_1}$ expansion using the
numerical values for the coefficients $y_n$'s entering in the expression (\ref{gdef}) and given in ref. \cite{Candelas}.

\section{PERTURBATIVE HETEROTIC DUALITY GROUP}

In this section, we wish to determine the perturbative duality group of
the present heterotic model in order to compair it with the analogous one
of the rigid theory of Seiberg and Witten. If the latter is realy the flat
limit of the low energy limit of the former, there should be some agreement
between the two groups.
 
As we already mentioned, at the classical level, the $T$-duality group of the
heterotic model is ${\Gamma_{0}(2)}_+$, which is  generated by the transformations
$W$ and $V$ defined in eq. (\ref{W}). These generators obey the
relations 
\begin{equation}
W^2=V^4=1 \qquad\qquad UVW=1 \ .
\label{group rel}
\end{equation}

The order 2 generator $W$ corresponds to the Weyl reflection
of the $SU(2)$ gauge group at the enhanced symmetry point $T=i/{\sqrt 2}$ \cite{Giveon}. 
Away from this point, $SU(2)$ is spontaneously broken by the
vacuum expectation value, $a$, of a Higgs field along the flat direction of the
scalar potential, with the identification $a\propto (T-i/{\sqrt 2})/(T+i/{\sqrt 2})$
near the non-abelian point. Thus, the ${\bf Z}_2$ transformation $T\rightarrow
WT=-1/(2T)$ acts on $a$ as the parity $a\rightarrow -a$ which remains unbroken
after the Higgs phenomenon.

At the quantum level, because of the singularity at $T=i/{\sqrt 2}$,
when moving a point around the singularity, the one loop prepotential $f$
transforms as:

\begin{eqnarray}
M_{i/\sqrt 2}: &f(T)& \rightarrow  f(T) + {\cal M}_{i/\sqrt 2}(T)   \\
{\cal M}_{i/\sqrt 2}(T) &=&
-4 \left( T-{i\over{\sqrt 2}} \right)^{2} \left( T+{i\over{\sqrt 2}} \right)^2.
\nonumber
\end{eqnarray}
The polynomial ${\cal M}_{i/\sqrt 2}$ can be determined by expanding the expression
(\ref{d5f}) around $T=i/{\sqrt 2}$. The T-duality group 
$\Gamma_0(2)_+$ is then modified at one loop level to a new one $G$ those 
relations are
\begin{equation}
W^2=M_{i/\sqrt 2}\qquad V^4=1 \qquad UVW=1 \ ,
\label{group qrel}
\end{equation}
where $M_{i/\sqrt 2}$ stands for the perturbative non-trivial monodromy.

One can now use the modified group relations (\ref{group qrel}) to determine the
transformation properties of $f$ under the generators $W$ and $V$. This
amounts to
determine the corresponding quartic polynomials ${\cal M}$ entering in
eq. (\ref{transfo modulaire}), which we will denote by ${\cal M}_W$ and
${\cal M}_V$ respectively. The result is the following:
\begin{equation}
{\cal M}_W = -4(1+w_{0})T^{4}+2w_{1}T^{3}-2T^{2}+w_{1}T+w_{0} \label{pol1}
\end{equation}
\begin{equation}
{\cal M}_V = \sum_{n=0}^4 v_n T^n
\end{equation}
with
\begin{equation}
v_0=1+w_0  \; , \, v_4=v_3-{4\over 3}v_2+2v_1-4v_0 \; , \label{pol2}
\end{equation}
\begin{equation}
{\cal M}_U = -{\cal M}_W(T+1) -4(T+1)^4{\cal M}_V\left({-1\over 2T+2}\right) \label{pol3} 
\end{equation}

The five independent parameters $w_{0,1}$ and $v_{1,2,3}$ entering in the
polynomials ${\cal M}_W$ and  ${\cal M}_V$ can be chosen arbitrarily using the
freedom to add to $f$ a quartic polynomial ${\cal P}(T)$ with real coefficients, and 
redefining the dilaton as
$S\rightarrow S+{\cal P}_{TT}/12$.

Finally, the full perturbative
symmetry group is the direct product of $G$ with the real constant dilaton shift,
\begin{equation}
D:\quad S\rightarrow S+\lambda \ .
\label{D}
\end{equation}

The symplectic structure of $N=2$ supergravity implies that all symmetry
transformations of the effective low energy theory must be contained in the
symplectic group $Sp(6,{\bf R})$ \cite{{sugra},{Ceresole}} which is broken to
$Sp(6,{\bf Z})$ by
quantum effects. It is then convenient to introduce a field basis where all
transformations act linearly. When we obtain a matrix representation of the
perturbative heterotic duality group, we can compair it to the
rigid case one. We define three homogeneous coordinates $X^I$, $I=0,1,2$ by
\begin{equation}
T=\frac{X^{1}}{X^{0}}\qquad \mbox{ and }\qquad S=\frac{X^{2}}{X^{0}}\ .
\label{hombas}
\end{equation}
In terms of these variables, the prepotential is a homogeneous function of degree 2
\begin{equation}
F(X^{0},X^{1},X^{2})=(X^{0})^{2} F\left(\frac{X^{1}}{X^{0}},\frac{X^{2}}{X^{0}}\right)\ ,
\end{equation}
and the K{\"a}hler potential takes the form
\begin{equation}
K = - \ln [ i ( \bar{X}^{I} F_{I}-X^{I} {\bar F}_{I} ) ] \; {\rm with} \; 
F_{I}=\frac{\partial F}{\partial X^{I}}\ .
\label{potentiell}
\end{equation}
In this way, all symmetries must act in the basis $(F_I,X^I)$ as symplectic
transformations which leave the K{\"a}hler potential (\ref{potentiell})
manifestly
invariant. Their symplectic action on the homogeneous basis is uniquely
defined from the corresponding transformations of the fields $T,S$ and the
prepotential $F$. Now, any choise of the parameters $\lambda_{W,V,U}$, $w_{0,1}$ and $v_{1,2,3}$ entering
eq. (\ref{Stransf}), (\ref{pol1}) and (\ref{pol2}) are allowed, but a suitable one can be
used to simplify the resulting representation of the generators $W$, $V$ and
$D$. In particular,
when written in some particular basis $(\widehat{F}_I,\widehat{X}^I)$, the matrix $\widehat{W}$ which represents the generator $W$ whose
classical
part corresponds to the Weyl reflection of the $SU(2)$ enhanced gauge
symmetry coincides with the perturbative
monodromy $M_{\infty}$ of the rigid theory \cite{Seiberg-Witten}:
\begin{equation}
\widehat{W} = {\widehat M}_{\infty}=    \left(  \begin{array}{cccccc}
                -1&0&0&2&0&0 \\
                0&1&0&0&0&0 \\
                0&0&1&0&0&0 \\
                0&0&0&-1&0&0 \\
                0&0&0&0&1&0 \\
                0&0&0&0&0&1
                \end{array}     \right) \;. \label{base}
\end{equation}
Near the enhanced symmetry point, ${\widehat X}^0\sim 2i(T-i/{\sqrt 2})$
and the 
subspace $({\widehat F}_0,{\widehat X}^0)$ is associated to the rigid
supersymmetric theory. The generator $\widehat{W}$ acts non-trivially only
on this
subspace and its action represented by the corresponding $2\times 2$
submatrix can
be identified with $M_{\infty}$ of ref. \cite{Seiberg-Witten}. This result
completes the check that the perturbative heterotic model considered here
is realy a string generalisation of the rigid model of Seiberg and Witten.

\section{EXACT TYPE II MONODROMY GROUP}

Our first task is to identify the generators of the perturbative heterotic
duality group derived in the previous section as elements of the type II monodromy group. The latter was worked out in ref. \cite{Candelas} and was shown
to be a subgroup of $Sp(6,{\bf Z})$, generated by 3 elements denoted by
$A$, $T$ and $B$.\footnote{To keep the same notation as ref.
\cite{Candelas}, we are forced to use the symbol $T$ for the generator which
should not be confused with the heterotic modulus.} The element $A$ generates
an exact ${\bf Z}_8$ symmetry and satisfies
$A^8=1$. The other two generators $T$ and $B$ are associated to the monodromies
around the conifold and the strong coupling loci, respectively, described by the
discriminant (\ref{conifold}), and they are subject to some group relations
given in ref. \cite{Candelas}.

Firstly, by considering the large complex structure limit, two
independent (mutually commuting) translations were identified, acting on the
special coordinates
$t_{1,2}$ of eq. (\ref{mirror map}),
\begin{eqnarray}
\label{tshifts}
S_1 &=& (AT)^{-2} \; : t_1\rightarrow t_1+1  \nonumber\\
S_2 &=& (ATB)^{-1} \; :\quad t_2\rightarrow t_2+1 \ .
\end{eqnarray}
{}From the identification with the heterotic variables $t_1=T$
and $t_2=S$, one concludes that in a suitable basis one should have:
\begin{equation}
S_1=U\qquad S_2=D|_{\lambda =1} \ .
\label{s12}
\end{equation}
The one loop heterotic
prepotential $f$ is shifted by the polynomial ${\cal M}_U$ of eq. (\ref{pol3})
under the action of $U$, while it remains inert under $D$. These transformations
allow us to determine the integration constants $P(t_1,t_2)$ entering in the expression of the
type II prepotential (\ref{crochet}) for which the identification (\ref{s12})
happens to be valid.

Secondly, the fact that the generator $A$ is of order 8 suggests that the order 4 generator
$V$ should be identified with a conjugation of $A^2$ (or its inverse). 
Indeed it
is easy to verify that the  matrix representation of $V$ and $A^2$ are identical.
Thus, we have shown that the
perturbative dualities generated by $U$ and $V$, together with the quantized
dilaton shift $D|_{\lambda=1}$, form a subgroup of the type II symmetries.
Using eqs. (\ref{tshifts}), (\ref{s12}) and the group relations (\ref{group qrel})
one has:
\begin{eqnarray}
U=(AT)^{-2} &,&  V=A^2 \; , \\
W=A^{-1}TAT &,& D|_{\lambda=1}=(ATB)^{-1} \; . \nonumber
\label{pertgr}
\end{eqnarray}
This is an additional perturbative check of the equivalence of the heterotic and type II
models considered here. 

Our next task is to give a non perturbative argument for this equivalence.
The heterotic model is not totaly unknown at the exact level since the flat limit
of its low energy limit is just the theory of Seiberg and Witten which is 
completly known. This non perturbative information happens to be equivalent
to the knowledge of the exact monodromy group in the rigid case \cite{Seiberg-Witten}.
Therefore,  our aim is now to identify the quantum monodromy group $\Gamma(2)$ of the
$SU(2)$ rigid field theory as a subgroup of the exact monodromy group of the 
type II theory. The rigid group is generated by two
elements, $M_{\infty}$ and $M_1$, which satisfy the relation \cite{Seiberg-Witten}:
\begin{equation}
M_{\infty}=M_1M_{-1}\ ,
\label{gamma2}
\end{equation}
where $M_{\infty}$ is the perturbative monodromy, while $M_1$ and $M_{-1}$
correspond to the monodromies around the points where dyonic hypermultiplets
become massless and they are conjugate to each other. These properties can be
used as a guide for the identification \cite{Lust}. We have shown in the
previous section that $M_{\infty}$ coincides with the generator $W$ of the
heterotic duality group. A simple inspection of its form (\ref{pertgr}) suggests
that $M_1$ should be identified with $T$ (or its conjugate $A^{-1}TA$). Indeed
one can easily verify that in the basis used in eq. (\ref{base}) the generator $T$
takes the form:
\begin{equation}
{\widehat T} ={\widehat M}_1=\left(     \begin{array}{cccccc}
                1&0&0&0&0&0 \\
                0&1&0&0&0&0 \\
                0&0&1&0&0&0 \\
                -2&0&0&1&0&0 \\
                0&0&0&0&1&0 \\
                0&0&0&0&0&1
                \end{array}     \right)\; .
\end{equation}
One sees that in analogy with ${\widehat W}$,  ${\widehat T}$ acts non-trivially
only on the subspace of the rigid supersymmetric theory $({\widehat F}_0,{\widehat
X}^0)$, and its action represented by the corresponding $2\times 2$ submatrix
coincides with $M_1$ of ref. \cite{Seiberg-Witten}. Thus, we have shown that the
non perturbative monodromies of the $SU(2)$ rigid field theory form a $\Gamma(2)$
subgroup of the type II monodromy group,
\begin{equation}
M_{\infty}=W,\; M_1=T, \; M_{-1} = (AT)^{-1}T(AT) \; .
\label{rigidgr}
\end{equation}

Finaly, let us note that the heterotic $T$-duality group extended with the quantized dilaton shift and the
non perturbative monodromy of the rigid field theory 
(\ref{rigidgr}) form a subgroup of the exact type II symmetries generated,
for instance, by the elements $A^2$, $A^{-1}TA$, $A^{-1}B$, and
$T$. One needs just to introduce the ``square root" of the perturbative generator
$V=A^2$, or equivalently $B$, to recover the full type II group. The generator $B$
is related to the monodromy around the non perturbative singularity $y=1/4$
corresponding to infinite coupling, where new massless dyonic
hypermultiplets appear with charges under the ``dilaton" $U(1)$. This singular
line which is not present in the rigid theory is a new stringy phenomenon related
to the dilaton and seems to be a generic feature of string vacua. From the matrix representation of the generator $B$, one finds
that it corresponds to the transformation,
\begin{equation}
T\leftrightarrow S'
\end{equation}
where $S'\equiv S+T-1$ should be the
correct identification of the heterotic dilaton based on the physical requirement
that the transformation $B$ preserves the positivity of the imaginary part of
the dilaton, {\em i.e.} of the inverse square of the coupling constant \cite{Klemm.95}.

\end{document}